\documentclass[aps,prl,floatfix,twocolumn,superscriptaddress]{revtex4}
\usepackage{latexsym}
\usepackage{graphicx}
\usepackage{rotating}
\usepackage[normalem]{ulem}
\usepackage{hyperref}
\usepackage{amsmath,amssymb,amsfonts}
\usepackage{bm}
\usepackage{comment}
\usepackage{color}
\usepackage{amsfonts}
\usepackage{mathrsfs}
\newcommand{\beq}{\begin{equation}}
\newcommand{\eeq}{\end{equation}}

\newcommand{\srvo}{SrVO$_{3}$}

\newcommand{\lyxmathsym}[1]{\ifmmode\begingroup\def\b@ld{bold}
  \text{\ifx\math@version\b@ld\bfseries\fi#1}\endgroup\else#1\fi}

\begin{document}

\title{A novel continuous time quantum Monte Carlo solver for dynamical mean field theory in the compact Legendre representation}

\author{Evan Sheridan}
\author{C\'edric Weber}
\author{Evgeny Plekhanov}
\affiliation{ Theory and Simulation of Condensed Matter, Department of Physics,
              King's College London, The Strand, London WC2R 2LS, UK. }
              
\author{Christopher Rhodes$^{*}$}
\affiliation{ Theory and Simulation of Condensed Matter, Department of Physics,
              King's College London, The Strand, London WC2R 2LS, UK. {\&} AWE,
              Aldermaston, Reading, RG7 4PR, UK. }

%%%%%%%%%%%%%%%%%%%%%%%%%%%%%%%%%%%%%%%%%%%%%%%%%%%%%%%%%%%%%%%%
%%%%%%%%%%%%%%%%%%%%%%%%%%%%%%%%%%%%%%%%%%%%%%%%%%%%%%%%%%%%%%%%
%%%%%%%%%%%%%%%%%%%%%%%%%%%%%%%%%%%%%%%%%%%%%%%%%%%%%%%%%%%%%%%%
%%%%%%%%%%%%%%%%%%%%%%%%%%%%%%%%%%%%%%%%%%%%%%%%%%%%%%%%%%%%%%%%
%%%%%%%%%%%%%%%%%%%%%%%%%%%%%%%%%%%%%%%%%%%%%%%%%%%%%%%%%%%%%%%%

\begin{abstract}
  Dynamical mean-field theory (DMFT) is one of the most widely-used methods to treat accurately electron correlation effects in \textit{ab-initio} real material calculations. Many modern large-scale implementations of DMFT in electronic structure codes involve solving a quantum impurity model with a Continuous-Time Quantum Monte Carlo (CT-QMC) solver \cite{CT-QMC_1,CT-QMC_2,CT-QMC_3,CT-QMC_4}. The main advantage of CT-QMC is that, unlike standard quantum Monte Carlo approaches, it is able to generate the local Green's functions $G(\tau)$ of the correlated system on an arbitrarily fine imaginary time $\tau$ grid, and is free of any systematic errors. In this work, we extend a hybrid QMC solver proposed by Khatami \textit{et al.} \cite{khatami} and Rost \textit{et al.} \cite{rost} to a multi-orbital context. This has the advantage of enabling impurity solver QMC calculations to scale linearly with inverse temperature, $\beta$, and permit its application to $d$ and $f$ band materials. In addition, we present a novel Green's function processing scheme which generates accurate quasi-continuous imaginary time solutions of the impurity problem which overcome errors inherent to standard QMC approaches. This solver and processing scheme are incorporated into a full DFT+DMFT calculation using the CASTEP DFT code \cite{CASTEP}. Benchmark calculations for {\srvo} properties are presented.
\end{abstract}

\maketitle

%%%%%%%%%%%%%%%%%%%% END OF ABSTRACT %%%%%%%%%%%%%%%%%%%%

%%%%%  MAIN TEXT -----

\section{I. Introduction}

Density functional theory (DFT), in both its local density (LDA) and generalised gradient (GGA) approximations, is a highly effective method for the calculation of quantitatively accurate \textit{ab-initio} ground-state properties of a wide range of real materials \cite{DFT-ref}. However, for some materials DFT's outstanding predictive capabilities become significantly less reliable. Such materials are found typically amongst the transition metal oxides, as well as in lanthanide and actinide-based compounds. This is problematic because of the potential application of these materials in the fields of quantum computing, data storage and high temperature superconductivity. These materials characteristically have narrow bandwidths ($W$ $\sim 2-3$ eV), so the correlations induced by the Coulomb interaction between valence electrons ($U$ $\sim 4-5$ eV) are strong, implying that \textit{i.e.} $U/W \gtrsim 1$. When bandwidths are broader the ratio of electron correlation to bandwidth is reduced (\textit{i.e.} $U/W \lesssim 1$), so the effects of electron correlation are weaker. In the latter case, the typical approximations involved at the DFT level work well, often with a quantitative precision that sustains comparison with experiment.

A single unified framework is sought in which the effects of electron correlation can be addressed for real material calculations over a wide range of temperatures and correlation strengths $U/W$. In recent years, a combination of DFT and dynamical mean-field theory (DMFT) has proved to be an effective way of interpolating between the itinerant and strongly localised limits of the electronic behaviour. The DFT+DMFT approach has been well established, and allows extending the capabilities of DFT-based approaches into calculations where electron correlation effects are significant. DFT+DMFT has evolved into a powerful method for dealing with correlation effects, and is being frequently used for material-specific applications. In addition to DFT+DMFT, the GW+DMFT approach is continuing to mature into an effective way of undertaking material-specific calculations for correlated systems \cite{GW+DMFT}. 

The critical component of a DFT+DMFT calculation is the so-called ``impurity solver'', which represents the quantum many-body physics interactions of the correlated electrons in the material. It is essential that the solver contains an accurate representation of the physics of the interacting electrons, and can efficiently compute solutions in fast and stable way over a wide range of parameters ($U$, $W$ and $T$). A variety of solvers are available for use in DMFT calculations, and they can be selected to match the computational resources available to complete a calculation in a time-efficient way. These solvers can be based on quasi-analytical methods (\textit{e.g.} Hubbard I, Iterated Perturbation Theory (IPT), Non-Crossing Approximation (NCA) or fluctuation exchange approximation (FLEX)) or numerical methods (\textit{e.g.} Numerical Renormalisation Group (NRG), Exact Diagonalisation (ED), QMC) \cite{DMFT_review}. Of all these methods it is the quantum Monte Carlo methods which have proved to be particularly popular. This is because QMC methods are conceptually straightforward and can return stable results (to statistical accuracy) over a wide range of parameter values.

A QMC technique known as continuous-time QMC (CT-QMC) \cite{CT-QMC_1, CT-QMC_2, CT-QMC_3, CT-QMC_4} has emerged lately as a popular solver for many DMFT applications. The main advantage of CT-QMC is that it overcomes one of the most significant limitations of conventional QMC methods - namely, the systematic errors which arise due the Trotter-Suzuki decomposition and the concomitant discretisation of the imaginary time interval. Errors of this kind can generate substantial bias in the results of DMFT calculations. This is because they can significantly shift the converged fixed point of the DMFT iteration \cite{rost}. However, a disadvantage of most of the readily available CT-QMC solvers is that the computational effort scales cubically with inverse temperature $ \beta = 1/k_{B}T $, and this severely limits their access to low temperature phases. Although variants of CT-QMC have been proposed recently which scale linearly in $\beta$ \cite{CT-QMC_linear} the method is still a perturbative one, so it generates noisy results for low temperatures and large values of electron correlation unless exceptional computational resources are available.  For multi-orbital or cluster DMFT applications, the inverse temperature scaling limitation becomes even more significant. A key computational bottleneck in calculations of fully charge self-consistent DFT+DMFT properties of real materials is the requirement to generate accurate impurity model solutions on a fast time-scale with medium-scale computational resources. For such challenging applications QMC methods remain a competitive technique for solving quantum impurity problems. Also, when calculating a material equation of state, numerous DFT+DMFT total energy calculations are needed over a range of temperatures and pressures \cite{mehta}. To perform these calculations to a level comparable with experimental data demands accurate and computationally efficient impurity solvers \cite{erik, mcmahan}.    

In this paper we present a QMC solver which scales linearly in inverse temperature \cite{khatami, rost} extended to a multi-orbital context, and supplement the solver with a robust Lagrange/Chebyshev-based Green's function interpolation technique that facilitates controlled extrapolations of the impurity Green's functions on multiple discretised imaginary time grids to a solution that is quasi-continuous in imaginary time. Because the solver self-energy is very sensitively dependent upon this Green's function it is imperative that the impurity model solution is an accurate one. In this way we are able to calculate accurate quasi-continuous time quantum impurity model solutions for multi-orbital systems faster than CT-QMC solvers, particularly at temperatures those solvers find challenging to access. 

The paper is organised as follows. In Section II we introduce the QMC solver and describe how it is integrated into a DFT+DMFT scheme for real material calculations using the CASTEP plane-wave DFT code \cite{CASTEP}. In Section III a new technique designed to generate accurate quasi-CT Green's functions using fits of multiple low-resolution QMC solutions is presented. Some of the particular issues of fitting Green's functions from real materials are discussed. In Section IV, to test the technique fully, we undertake a full DFT+DMFT calculation on strontium vanadium oxide {\srvo} by integrating our solver with both the CASTEP DFT code and the quasi-CT Green's function scheme. Where appropriate, calculations are base-lined against equivalent quasi-CT QMC calculations using the TRIQS solver that is embedded in CASTEP \cite{CASTEP} \cite{evgeny_Ce}. All energies and temperatures are in eV.

%%%%%%%%%%%%%%%%%%%% END OF SECTION I %%%%%%%%%%%%%%%%%%

\section{II. QMC impurity solver and DFT+DMFT calculations}

\subsection{1. The QMC algorithm}

The purpose of a quantum impurity solver in a DMFT calculation is to adequately capture the physics of spin-dependent electron-electron interactions, using either a numerical model Hamiltonian representation or a quasi-analytic approach \cite{DMFT_review}. The initial conditions for the model are derived from a material-specific DFT calculation, and then the solver solution is processed and fed back to the DFT code in an iterative cycle. The key to efficient DFT+DMFT calculations of real material properties is finding a fast, accurate and stable solution to the impurity model (without a fermionic sign problem \cite{dos_santos}) over a wide range of model parameters ($U$, $J$ and $W$) and inverse temperatures, $\beta$.

Most early DFT+DMFT calculations employed a Hirsch-Fye (HF) solution of the single impurity Anderson model to represent electron correlation effects \cite{DMFT_review}. The appeal of this solver is that it is QMC-based method which can be applied over a broad parameter range and band fillings without being compromised by fermionic sign problems. However, this method contains a systematic error inherent in the Trotter decomposition and imaginary time discretisation. Also, its inverse temperature scaling goes as $\beta^{3}$. More recently HF-QMC has been complemented by the CT-QMC method (and variants thereof) discussed above. But, as noted there, it too scales as $\beta^{3}$. Another popular impurity solver is based on exact diagonalisation (ED) of the impurity Hamiltonian \cite{caffarel, cedric_ed}. The advantage of this method is that it gives an exact Green's function solution and scales linearly in $\beta$. However, it relies on a discretisation of the Weiss field (bath Green's function), and the computational load scales exponentially with the number of bath sites which can be limiting for applications beyond application to single site, single band, impurity models.

Khatami \textit{et al.} \cite{khatami} and later Rost \textit{et al.} \cite{rost} introduced a novel Hamiltonian-based QMC scheme that combined the virtues of ED with a standard QMC-type solution, resulting in an impurity solver that scales cubically in the number of discretised bath sites and linearly in inverse temperature. Moreover, it has the same fermion sign performance as HF-QMC. Rost \textit{et al.} \cite{rost} demonstrated the advantages of using this type of solver in DMFT studies of the single impurity model on a Bethe lattice \cite{DMFT_review}. The solver still exhibits the systematic Trotter errors characteristic of QMC methods, but, as we show below, these can be treated by the application of a novel Green's function interpolation and extrapolation technique which generates a quasi-CT solution.

The formulation of the QMC algorithm for single band physics is described in detail in \cite{khatami} and \cite{rost}. In what follows we will be extending the method in a multi-band context. For the general case of electrons on a single impurity site which interact both on and between $2M$ orbital/spin pairs (where $1 < m \leq 2M$), the Hamiltonian is given by
\begin{eqnarray}
H = H_{LDA}^{0} & + & \sum_{m}U_{mm}n_{m\uparrow}n_{m\downarrow} - \sum_{m \sigma} \mu n_{m \sigma} \nonumber \\ 
    & + & \sum_{m < m' \sigma} U_{m m'} n_{m \sigma} n_{m' -\sigma} \nonumber \\
    & + & \sum_{m < m' \sigma} (U_{m m'} - J_{m m'}) n_{m \sigma}n_{m' \sigma}
    \label{eq:mo_hamiltonian}  
\end{eqnarray}
where $U_{m m} = U$, $J_{m m'} = J$, and $U_{m m'} = U - 2J$ for $m \neq m'$. Also, $n_{m\sigma} = c^{\dagger}_{m\sigma}c_{m\sigma}$ where $c^{(\dagger)}_{m\sigma}$ are the annihilation (creation) operators for electrons in orbital $m$ with spin $\sigma$.  In this formulation there are Hubbard-like interactions between opposite-spin electrons in the same orbitals and neighbouring orbitals, and Ising-like Hund's coupling between same-spin electrons on neighbouring orbitals. The full Slater-Kanamori Hamiltonian for multi-orbital interactions is rotationally invariant in spin-space \cite{sakai} and also includes ``spin-flip'' and ``pair-hopping'' terms, which are likewise determined by $\approx J_{m m'}$. These interactions cannot be straightforwardly factorised into products of quadratic operators $n_{m\sigma}$, which is essential in any QMC-based implementation. Various schemes have been proposed to include these additional terms into QMC formulations \cite{sakai, sakai_2004,  han,  motome_I,  motome_II,  held_vollhardt}, but this is not wholly straightforward to do without introducing additional sign problems, especially at low temperatures. We do not include spin-flip and pair-hopping terms in this work. ED and CT-QMC approaches can handle the full Hamiltonian, and this must be borne in mind when comparing results generated by different multi-orbital impurity solvers.   

Our solver uses a multi-orbital generalisation of the standard determinantal QMC procedure. For a multi-orbital interaction matrix $\sum_{m m'} U_{m m'} n_{m} n_{m'}$ the usual Hubbard-Stratonovich transformation can be applied, \textit{i.e.}

\begin{eqnarray}
& & e^{-U_{m m'}\Delta\tau (n_{m}-1/2)(n_{m'}-1/2)} = \nonumber \\
& &\frac{1}{2} \sum_{s m m' = \pm 1} e^{\lambda_{m m'}s_{m m'}} (n_{m} - n_{m'})
\label{eq:trotter}
\end{eqnarray}
where $\lambda_{m m'} = {\text {acosh}}[e^{(\frac{1}{2} \Delta\tau U_{m m'})}]$ and $U_{m m'}$ is one of $U$, $(U-2J)$ or $(U-3J)$, depending on which orbital/spin pair is being considered. There is an auxiliary spin fields $s$ for each inter and intra orbital interaction.

Following conventional QMC procedure, at each imaginary time slice each of the $M(2M-1)$ auxiliary spin fields is flipped sequentially and then tested for acceptance or rejection using the ratio $R = R_{m}R_{m'}$ where

\begin{eqnarray}
& & R_{m}  = 1 + \{ (1 - G_{m}) \gamma_{m} \} \nonumber \\
& & R_{m'} = 1 + \{ (1 - G_{m'}) \gamma_{m'} \}
\label{eq:ratio}
\end{eqnarray}
with $\gamma_{m} = e^{2 \lambda_{m m'}s_{m m'}-1}$ and $\gamma_{m'} = e^{-2 \lambda_{m m'}s_{m m'}-1}$ . If the spin flip is accepted the Green's function is updated according to the standard BSS \cite{BSS} prescription
\begin{eqnarray}
& & G_{j,k,m} = G_{j,k,m} - (\delta_{j,1} - G_{j,1,m})\gamma_{m}G_{1,k,m}/R_{m} \nonumber \\
& & \hspace{-0.6cm} G_{j,k,m'} = G_{j,k,m'} - (\delta_{j,1} - G_{j,1,m'})\gamma_{m'}G_{1,k,m'}/R_{m'}.
\label{eq:gf_update}
\end{eqnarray} 

If the spin flip is rejected the Green's functions remain the same. The condition for half-filling in this Hamiltonian is given by $\mu = (U + (M-1)(U-2J) + (M-1)(U-3J))/2$.

In the exact diagonalisation method \cite{caffarel} the infinite lattice surrounding the impurity site is approximated by a discretised bath of finite size $N_{b}$, and the identical procedure is followed in this QMC solver. The first step is to parameterise the Weiss field ${\cal{G}}_{0}$ for each orbital and spin in terms of a finite number of bath parameters by approximating the Weiss Green's function in terms of a non-interacting Anderson impurity model as follows:
\begin{equation}
{\cal{G}}_{And,m}^{-1}(i\omega_{n}) = i\omega_{n} + \mu - \epsilon^{imp}_{m} - \sum_{k=1}^{k=N_{b}} \frac{V_{m,k}V_{m,k}^{*}}{i\omega_{n} - \epsilon_{m,k}}
\label{eq:gf_and}
\end{equation} 
where $k$ is the index for the bath level, $m$ is the index for each orbital/spin. Essentially, this entails minimising the difference between the Weiss field and equation (\ref{eq:gf_and}) using a cost function, typically
\begin{equation}
\chi^{2}[{\epsilon_{k}, V_{k}}] = \sum_{n=0}^{n=n_{c}} {\cal{A}}_{n} | {\cal{G}}_{And}(i\omega_{n}:\{\epsilon_{k}, V_{k}\}) - {\cal{G}}_{0}(i\omega_{n}) |^{2}. 
\label{eq:chi_sq}
\end{equation}
In practice it has been found advisable to weight the cost function towards smaller imaginary frequencies by using  pre-factor ${\cal{A}}_{n} \approx 1/ i\omega_{n}^{2}$. This avoids squandering fitting cost on the asymptotic regions of the Green's function, and has been found particularly apposite when attempting accurate fits to real material Green's functions. A robust multi-dimensional fitting routine was used to determine the set of $\epsilon_{k}$ and $V_{k}$ for each orbital, which are subsequently used to parameterise the QMC solver. To avoid falling into local minima, multiple fits over a wide range of initial conditions are performed and the best-fit is then selected.   

Once a set of $\epsilon_{k}, V_{k}$ has been found to fit the Anderson impurity Green's function the QMC part of the calculation proceeds as a standard Monte Carlo simulation for a multi-orbital context, where the imaginary time interval $0 \leq \tau \leq \beta $ is discretised into $L$ time-slices ($\beta = L\Delta\tau $). A sufficient number of Monte Carlo steps are run on a parallelised processor configuration, and the imaginary time impurity Green's functions are calculated for each orbital and spin using the procedure of White \textit{et al.} \cite{white}, and Loh and Gubernatis \cite{loh+gub}.

\subsection{2. DFT+DMFT implementation with CASTEP}

In our implementation of DFT+DMFT the plane-wave DFT code CASTEP was used \cite{CASTEP}. It is a widely available DFT code that can be readily used for calculating \textit{ab-initio} material properties. Previous work with CASTEP has demonstrated the integration of a Hubbard I solver \cite{evgeny_Ce} and an ED-based solver \cite{cedric_ed} for material-specific calculations, and what follows builds on that. The multi-orbital solver described above was integrated with CASTEP to better represent the electron correlation effects. Many of the issues associated with integrating an impurity solver with a DFT code are generic and a variety of methods have been used. Here we give a description of our CASTEP implementation of DFT+DMFT, and this can be compared to other approaches.

A CASTEP calculation generates an LCAO basis set using either norm-conserving or ultra-soft pseudopotentials. These two cases can be dealt with on an equal footing by defining an ``overlap matrix'', $S$. The basis set functions generated from norm-conserving pseudopotentials are orthogonal by construction, so the overlap matrix is the identity matrix. In the case of ultra-soft pseudopotentials these states are overlapping with a matrix:
\begin{equation}
\langle \chi_{m' \bf R'} |S| \chi_{m \bf R} \rangle = \delta_{m',m}\delta_{\bf R,\bf R'}.
\end{equation}        
This implies that the Kohn-Sham (KS) equations transform from a standard eigenvalue problem into a generalised one
\begin{equation}
H_{\bf k}^{KS} | \Psi_{{\bf k},\nu} \rangle = E_{{\bf k},\nu}S | \Psi_{{\bf k},\nu} \rangle
\end{equation} 
where $\Psi_{{\bf k},\nu}$ are the KS eigenstates.

\subsection{3. Projectors}

The standard DMFT technique is defined in terms of completely localised electronic states (\textit{e.g.} as in Hubbard model). On the other hand, in a wide class of {\it ab-initio} codes, including 
CASTEP, the electrons are described in terms of completely delocalised plane wave states. 
Consequently, a key feature of the DFT+DMFT calculation is the selection of a correlated sub-space of orbitals, and the means to bridge the DFT Bloch space basis and the basis of the correlated sub-space. To do this we define the orthonormal projectors $P_{L, \nu}({\bf k})$, where
\begin{equation}
P_{L,\nu}({\bf k}) = \langle \chi_{L} | S | \Psi_{{\bf k}, \nu} \rangle. 
\end{equation}
We now have two distinct spaces - \textit{i}) the KS Bloch space indexed by $k$ and $\nu$, and \textit{ii}) the localised (correlated) subspace indexed by $L$.

To go from $\chi_{L}$ to $\Psi_{{\bf k}, \nu}$ \textit{i.e.} ``up-folding'' , we use
\begin{equation}
| a_{{\bf k}, \nu} \rangle = \sum_{L} P_{\nu, L}^{*}({\bf k}) | b_{L} \rangle .
\end{equation}
Conversely, to go from $\Psi_{{\bf k}, \nu}$ to $\chi_{L}$ \textit{i.e.} ``down-folding'' , we use
\begin{equation}
| b_{L} \rangle = \sum_{{\bf k}, \nu} P_{L,\nu}({\bf k}) | a_{{\bf k}, \nu} \rangle,
\end{equation} 
where $a_{{\bf k},\nu}$ is a vector defined in Bloch space and $b_{L}$ is a vector defined in the space of correlated orbitals.

The projector matrix satisfies the following condition:
\begin{equation}
\sum_{{\bf k}, \nu} P_{L, \nu}({\bf k})P_{\nu, L'}^{*}({\bf k}) = \delta_{L,L'}.
\end{equation}   
In this scheme an up-folding operation followed by a down-folding operation is equivalent to an identity operation. Both the up-folding and down-folding operations are each used once in each DMFT iteration cycle, as we now discuss.

\subsection{4. The DMFT cycle}

To begin the DMFT iteration for {\srvo} we use CASTEP to calculate the Bloch Green's function 
\begin{equation}
G^{B}_{\nu,\nu'}({\bf k},i\omega_{n}) = \{ (i\omega_{n} + \mu - \epsilon_{{\bf k},\nu})\delta_{\nu,\nu'} - \Sigma^{B}_{\nu,\nu'}({\bf k},i\omega_{n}) \}^{-1} 
\label{eq:gf_bloch}
\end{equation}  
which is the Fourier transform $ F.T. [ \langle {\bf r} | \hat{G} | {\bf r} \rangle  ] $, where
\begin{equation}
\hat{G}({\bf r}, i\omega_{n}) = (i\omega_{n} + \mu + \frac{1}{2}\nabla^{2} - \nu_{KS}({\bf r}) - \Sigma^{B}({\bf r}, i\omega_{n}) )^{-1}.
\end{equation}
(An adjustment to the chemical potential $\mu$ is necessary at this point in the cycle to ensure correct level of electron occupancy - see Appendix A.) 
We now consider a correlated atom at location ${\bf R}$. The basis functions in the correlated sub-space are denoted by $m$. The local Green's function for the correlated sites is now obtained from the Bloch Green's function by down-folding and then averaging over the Brillouin zone, as follows:
\begin{equation}
G_{m,m'}^{loc} (i\omega_{n})  = \frac{1}{N_{{\bf k}}}\sum_{\nu, \nu', {\bf k}} P_{m,\nu}({\bf k}) G_{\nu,\nu'}^{B}({\bf k}, i\omega_{n}) P_{\nu',m'}^{*}({\bf k}).
\label{eq:gf_loc}
\end{equation}
We now make the identification of $G^{loc}$ with $G^{imp}$ - so in the DMFT impurity Dyson equation we calculate the Weiss field 
\begin{equation}
[{\cal{G}}_{0}(i\omega_{n})]^{-1}_{m,m'} = \Sigma_{m,m'}^{imp}(i\omega_{n}) + [G^{imp}(i\omega_{n})]_{m,m'}^{-1}.
\label{eq:gf_bath}
\end{equation}
On the first iteration we make a guess for the self-energy (typically $\Sigma^{imp} = 0$). We then parameterise the Weiss field $[{\cal{G}}_{0}^{-1}]_{m,m'}$ using equation (\ref{eq:gf_and}) for ${\cal{G}}_{And}^{-1} $. 

It is now possible to run the multi-orbital QMC solver (described above) to obtain a set of impurity Green's functions ($G_{QMC}(\tau)$).  Following the interpolation and extrapolation procedure (described in Section III), the Fourier Transformed Green's functions are used to calculate a revised impurity self-energy, \textit{i.e.}
\begin{equation}
\Sigma^{imp}_{m,m'} = [{\cal{G}}_{0}(i\omega_{n})]^{-1}_{m,m'} - [G_{QMC}(i\omega_{n})]^{-1}_{m,m'}.
\label{eq:se_loc}
\end{equation}
At this point it is necessary to make allowance for the ``double counting'' term, $V^{DC}_{m,m'}$ (see Appendix B), and then up-fold the impurity self-energy back to Bloch space , \textit{i.e.}
\begin{equation}
\Sigma^{B}_{\nu, \nu'}({\bf k}, i\omega_{n}) = \sum_{m,m'} P_{\nu,m}^{*}({\bf k})(\Sigma_{m,m'}^{imp}(i\omega_{n}) - V^{DC}_{m,m'}) P_{m',\nu'}({\bf k}).
\label{eq:se_bloch}
\end{equation} 
The self-energy is purely local when expressed in the set of correlated orbitals, but it acquires momentum dependence when up-folded to the Bloch basis set. The up-folded self-energy is returned to equation (\ref{eq:gf_bloch}), and the iteration sequence is continued until an acceptable level of convergence is reached in the self-energy (or chemical potential).

System properties can also be calculated during the DMFT iteration cycle. Of particular value are the Bloch-level occupation matrix:
\begin{equation}
N_{e} = \frac{1}{N_{\bf k}}\sum_{\nu,{\bf k}}\sum_{n}G^{B}_{\nu,\nu'}({\bf k},i\omega_{n})e^{i\omega_{n}0^{+}},
\label{eq:occ_bloch}
\end{equation}
the spectral density 
\begin{equation}
A_{\nu,\nu'}({\bf k},\omega) = -\frac{1}{\pi} Im G^{B}_{\nu,\nu'}({\bf k},\omega),
\label{eq:spec_dens}
\end{equation}
and the density of states (DOS)
\begin{equation}
D(\omega) = \frac{1}{N_{\bf k}} \sum_{\nu,{\bf k}} A_{\nu,\nu'}({\bf k},\omega).
\label{eq:dos}
\end{equation}
To calculate the spectral density and DOS, the analytic continuation from imaginary ($i\omega_{n}$) to real ($\omega$) frequency is made using the Pad\'e approximation \cite{vidberg}. 

As well as our implementation here, which uses the BSS method to update
the Green's function, a version of the multi-orbital solver code has also been 
implemented which uses the Nukala {\it et al.} \cite{nukala} fast updating 
method. The results are identical to BSS, but the Nukala method will be better 
suited to multi-orbital cluster DMFT applications.

%%%%%%%%%%%%%%%%%% END OF SECTION II %%%%%%%%%%%%%%%%%%% 

\section{III. A novel quasi-continuous imaginary time solution}

The conventional QMC method calculates imaginary time Green's functions $G(\tau)$ on a grid of ($L$) regularly spaced $\tau$ points over the interval $0 \leq \tau \leq \beta (= L\Delta\tau)$. In DMFT calculations this Green's function is Fourier transformed and then used to calculate the self-energy of the impurity problem. However, any attempt to calculate the frequency-dependent self-energy $\Sigma(i\omega_{n})$ using the raw QMC Green's functions immediately generates two problems, both related to the fact that the Green's function is calculated on a discrete imaginary time grid. Firstly, the Trotter decomposition implies that the resulting self-energy will be in error - to first order $\sim U \Delta \tau^{2}$. Secondly, attempting to Fourier transform the Green's function $G_{QMC}(\tau)$ as it stands will introduce aliasing errors in the transformed function $G_{QMC}(i\omega_{n})$. Aliasing arises from the fact that the Fourier transform of the imaginary time Green's function is a periodic function in imaginary frequency with poles at multiples of the Nyquist frequency ($2\pi L/\beta$). This leads to weight being folded back into the transform from frequencies above the Nyquist frequency, and generates incorrect asymptotic behaviour of $G_{QMC}(i\omega_{n})$.

In this section we present a novel multi-scale Green's function processing technique that simultaneously addresses these two issues. The method interpolates a set of comparatively coarse time-scale QMC impurity Green's functions on to a fine-scale $\tau$ grid (to eliminate aliasing problems), and extrapolates them to a generate a quasi-continuous imaginary time solution for the multi-orbital impurity problem (to eliminate systematic Trotter errors). This method generates Green's functions and self-energies with the correct behaviour across all imaginary frequencies that can be seamlessly integrated into the DMFT iteration cycle. 

Splining procedures (with and without quasi-CT extrapolation) have been used previously to process imaginary time Green's functions so they can be used in DMFT calculations \cite{rost}. In real material calculations the imaginary time Green's function can substantially change its magnitude over the space of few imaginary time intervals, $\Delta\tau$. For strong electron interactions and low temperature, this behaviour can be particularly acute. Because of this care must be taken when interpolating and extrapolating Green's functions so as not to inadvertently incorporate spurious processing pathologies into calculated Green's functions. Moreover, QMC methods generate noisy data which can cause additional challenges for splining schemes, so here we describe a new Green's function splining and extrapolation scheme which is robust and can fit noisy data before generating a quasi-continuous time solution to the DMFT Hamiltonian. As we will show subsequently, it is straightforward to integrate this scheme into a
full material-specific DFT+DMFT calculation.

\subsection{1. Polynomial basis method}

One way to reduce the systematic error introduced by the Trotter decomposition 
is to simply reduce $\Delta \tau$ by increasing the number of imaginary time 
steps $L$. However, the price for an increasingly accurate representation of 
$G(\tau)$ is a significantly more computationally intensive QMC calculation. For
example, the ubiquitous Hirsch-Fye solver scales as $\sim L^{3}$.

The objective of our new Green's function processing protocol is to demonstrate that by using a novel polynomial basis interpolation for $G_{QMC}(\tau)$, and an extrapolation scheme, we can perform QMC calculations on multiple, relatively coarse, imaginary time grids to generate a quasi-continuous imaginary time solution. 

We expand $G(\tau)$ in an arbitrary orthogonal polynomial basis $P_i^{(k)}[x(\tau)]$ (\textit{e.g.} Legendre, Chebyshev, etc) where $i$ is the polynomial order, $k$ is the polynomial species and $x(\tau) = \frac{2 \tau}{\beta} - 1$ is the transformation from $[-1,+1]$ to $[0,\beta]$. The expansion is:
\begin{equation}
G^{(k)}(\tau) = \sum_{i \geq 0} P_i^{(k)}[x(\tau)] G_{i}^{(k)}.
  \label{eq:g_tau}
\end{equation}
To isolate the basis coefficients we apply the orthogonality constraints
obeyed by the polynomials, \textit{i.e.}, 
\begin{eqnarray}
& & \int_{0}^{\beta}  d \tau G^{(k)}(\tau) P_{i'}^{(k)}[x(\tau)] W(x(\tau)) = \nonumber \\
& &\int_{0}^{\beta} d \tau P_{i}^{(k)}[x(\tau)] P_{i'}^{(k)}[x(\tau)] W(x(\tau)) G_{i}^{(k)}.
\end{eqnarray}
The general orthogonality condition obeyed by the family of polynomials $P_{i}^{(k)}[x(\tau)]$ is 
\begin{equation}
\int_{0}^{\beta} d \tau P_{i}^{(k)}[x(\tau)] P_{i'}^{(k)}[x(\tau)] W(x(\tau)) =
\tilde{W}(i) \delta_{i,i'}, 
\end{equation}
and so the basis coefficients can be calculated as
\begin{equation}
G_{i}^{(k)} = \frac{1}{\tilde{W}(\tau)} \int_{0}^{\beta}  d \tau G^{(k)}(\tau) P_{i}^{(k)}[x(\tau)] W(x(\tau)).
  \label{eq:g_coeffs}
\end{equation}

Here we restrict our analysis only to the Legendre polynomials, where
$W(\tau) = 1$ and $\tilde{W}(i) = \frac{1}{2i + 1}$.

Calculating $G_i$, allows us to express $G(\tau)$ on an arbitrarily fine 
imaginary time grid. 

It is imperative to first obtain a reliable representation of $G(\tau)$ in the Legendre basis. To achieve this, it is possible to formulate a controlled fitting procedure that uses the Legendre basis coefficients $g_l$ as parameters to be adjusted to the raw QMC data.

\subsection{2. Green's function fitting procedure}

Expressing $G(\tau)$ in the Legendre basis allows a fitting procedure
to be formulated, which amounts to the minimisation of the function
\begin{equation}
\text{min}_{\{ g_l \}} [ G_{QMC}(\tau) - G_{FIT}^{\{ g_l \}}(\tau)],
\label{eq:min_leg}
\end{equation}
where the fitted (model) function $G_{FIT}^{\{ g_l \}}(\tau)$ is parametrised by the basis coefficients $g_l$, {\it i.e.} 
\begin{equation}
G_{FIT}^{\{ g_l \} }(\tau) = \sum_{l \geq 0}^{N_l} \frac{\sqrt{2l +1}}{ \beta} P_{l}(x(\tau))  g_l .
  \label{eq:leg_fit}
\end{equation}
It is straightforward to find $G_{FIT}^{\{ G_l \} }(\tau)$ using the conjugate 
gradient method, and since $N_l$ (the number of Legendre coefficients) is
generally quite modest. In our case because $N_l \approx 20$, this
procedure is exceptionally computationally efficient.
Moreover, by shifting the paradigm of dealing with a statistical problem, {\it i.e.}
Monte Carlo, to that of an optimisation one, allows the advantage of 
including \emph{a-priori} information. This strategy becomes
particularly attractive when dealing with unrefined QMC data on
exceedingly coarse imaginary time grids. In this respect, the Legendre 
basis proves to be an extremely useful tool for two outstanding reasons; the first is its simple relationship to the moments of the $G_{QMC}(i \omega_n)$, and the second is due to the convergence properties of $g_l$ in the Kernel Polynomial
Method (KPM) \cite{KPM_ref}. The final significant constraint that can
be imposed on the parameters $g_l$ in an \emph{a-priori} fashion is the
convexity of $G(\tau)$. We now discuss how incorporating this additional
information on $G_{FIT}(\tau)$ can reliably improve its accuracy.

In many-body calculations it is essential that the Green's function solutions 
have the correct high frequency tail. Often the high frequency tail, which 
has a $1 / i \omega_n$ behaviour, is fitted onto the low frequency result. However, 
in the Legendre basis there is an exact relationship between the
moments of the Green's function and the Legendre coefficients \cite{Green_Lagrange}. 
By introducing a set of Lagrange parameter penalty terms into equation (\ref{eq:min_leg}) information on the tail can be included into the fit, thereby eliminating the need for an {\it ad-hoc} fit of the tail. The term added 
to equation (\ref{eq:min_leg}) to ensure the correct high frequency tail asymptotic behaviour is
\begin{eqnarray}
  \lambda_{1} & &\left(c_{1}  + \sum_{l \geq 0 , \text{even}} \frac{2 \sqrt{2l +1}}{\beta} g_l\right) \nonumber \\
  & + & \lambda_{2}\left(c_{2} - \sum_{l \geq 0 , \text{odd}} \frac{2 \sqrt{2l +1}}{\beta^2} g_l l(l+1)\right)
  \label{eq:leg_tail}
\end{eqnarray}

where $\lambda_1$ is the Lagrange parameter controlling the first moment $c_{1}$, 
and $\lambda_2$ is the Lagrange parameter controlling the second moment $c_{2}$. 
The inclusion of the two Lagrange penalty terms relies on knowing the values
for $c_{1}$ and $c_{2}$ for $G(i \omega_n)$ before the fit is performed. Fortunately, for $c_{1}$ it is known that in the high frequency limit of $G_{QMC}(i\omega_n)$,
it behaves as $1/(i\omega_n)$, and therefore $c_{1} = 1$. The equivalent considerations for $c_{2}$ are somewhat more involved, but it can be shown that
\begin{equation}
c_2 = \mu - \varepsilon + \Sigma'(\infty), 
  \label{eq:second_moment}
\end{equation}
where $\mu$ is the chemical potential, $\varepsilon$ is the impurity level and
$\Sigma'(\infty)$ is the high frequency real asymptotic self-energy of 
an isolated impurity. $\Sigma'(\infty)$ is attainable by solving the
Anderson Impurity Model with a finite set of bath orbitals; for simplicity,
this can be achieved using an ED-solver \cite{cedric_ed}, or Hubbard I solver 
\cite{DMFT_review}.

It is an unavoidable fact that the truncation of a function in any
polynomial basis that is, in principle, infinite can lead to convergence issues.
In the case where there are non-differentiable points or singularities it is
especially problematic and can lead to the well established Gibbs oscillations. The severity of the oscillations near these ill-defined points can 
be damped by the introduction of a kernel $k_l$ in equation
(\ref{eq:leg_fit}) such that $g_l \rightarrow k_l g_l$. The process of truncating this series and modifying the basis coefficients amounts convolving $G(\tau)$ with
a kernel $k_l$. Since $G(\tau)$ is continuously differentiable (except at its
boundaries) it is possible to pick a kernel that will guarantee this behaviour, 
and in the process filter out any spurious noise. In this work, we concern 
ourselves only with the kernels of Dirichlet and Jackson.

\subsection{3. Green's function extrapolation for $\Delta \tau^{2}$ scaling}

Representing $G(\tau)$ in a polynomial basis is the first stage in the 
two-step quasi-continuous method by generating an accurate parameterisation
of the raw QMC Green's functions on an imaginary time grid of arbitrary
resolution. The second step is the systematic removal of the 
Trotter error by engineering a well defined extrapolation procedure of the Legendre basis coefficients on different Green's functions 
$G_{\lambda}(\tau)$, where $\lambda$ is the imaginary time grid index, related to
$\Delta \tau_{\lambda} =  \beta / N_{\lambda}$, with $N_{\lambda}$ being
the number of imaginary time points. 

The procedure begins with defining a measure of error on each grid and time point,
{\it i.e.} $F^{\lambda}(\tau_i)$ such that, 

\begin{equation}
  F^{\lambda}(\tau_i)=  \sum_{l=1}^{N_p} [g_{l}^{\lambda} - \mathcal{T}_{l}] P_{l}[x(\tau_{i})] = \alpha(\tau_i) \Delta \tau_{\lambda}^{2}, 
  \label{eq:extrap}
\end{equation}

where $\mathcal{T}_{l}$ are Legendre basis coefficients of quasi-continuous 
Green's function $G_{\text{QC-QMC}}(\tau)$, {\it } absent of the systematic 
Trotter error. The $\alpha(\tau_i)$ are the scaling coefficients of each
imaginary time point. 

The motivation for defining such an object is that we would like to
find the set of coefficients $\{ g_{l}^{\lambda_{\text{QC}}}\}$ such that
$F^{\lambda_{\text{QC}}}(\tau_i)=0$. To find this set of coefficients 
we define the following minimisation problem:

\begin{equation}
  \text{min}_{\mathcal{T}_{l}, \alpha(\tau_i)} \left[   F^{\lambda}(\tau_i) - \alpha(\tau_i) \Delta \tau^2 \right],
\label{eq:min_extrap}
\end{equation}

over the completely continuous target parameters $\mathcal{T}_l$ and scaling
coefficients $\alpha(\tau_i)$. It is possible to simplify this procedure by
introducing the matrix $A_{\lambda}(\tau_i)$ such that,

\begin{equation}
  A_{\lambda}(\tau_i) = \sum_{l=1}^{N_p} \frac{[g_{l}^{\lambda} - \mathcal{T}_{l}]}{\Delta \tau_{i}^{2}} = \alpha(\tau_i), 
  \label{eq:a_matrix}
\end{equation}

and noticing that across each grid $\lambda$ that $A_{\lambda}(\tau_i)$ remains
unchanged. As a result, it is possible to map the minimisation problem of
equation (\ref{eq:min_extrap}) to that of a simpler one,

\begin{equation}
  \text{min}_{\mathcal{T}_{l}} \left| \frac{A_{\lambda}(\tau_i)}{A_{\lambda-1}(\tau_i)} - 1 \right|,
  \label{eq:min_extrap_a_matrix}
\end{equation}

that is simply a minimisation with respect to the target parameters $\mathcal{T}_l$
and not the scaling coefficients $\alpha(\tau_i)$. 

Of critical importance in generating a controlled estimate of the high frequency 
tails of $G_{\text{QC-QMC}}(i \omega_n)$, and thus also 
$\Sigma_{\text{QC-QMC}}(i \omega_n)$, is the implementation of the additional constraint on the second moment of $G_{\text{QC-QMC}}(i \omega_n)$, as discussed earlier. In principle, it is only sensible to add this constraint here, at the level of the  Green's function without any systematic error, since $G_{\text{QC-QMC}}(i \omega_n)$ truly represents a physically relevant object while the $G_{\lambda}(i \omega_n)$ are inherently non-physical.

%%%%%%%%%%%%%%%%%% END OF SECTION III %%%%%%%%%%%%%%%%%

\section{IV. Application to SrVO$_{3}$}

\subsection{1. DFT+DMFT for SrVO$_{3}$}

In this section we integrate the three main components of our DFT+DMFT scheme: the CASTEP DFT code, the multi-orbital QMC-based solver, and the quasi-continuous imaginary time Green's function processing protocol. 

We apply this scheme to a full iterated DFT+DMFT calculation of {\srvo} properties, and, where relevant, show comparison calculations made using the TRIQS CT-QMC solver.  

The lack of any structural or magnetic phase transition behaviour over a broad temperature range makes the metallic transition metal oxide {\srvo} an ideal test material against which to benchmark our CASTEP first-principles DFT+DMFT technique. There have been previous experimental and theoretical investigations of this material (using a variety of flavours of DFT+DMFT) against which results using our DFT+DMFT scheme can be compared \cite{anisimov_srvo3, nekrasov_srvo3, lechermann_srvo3, amadon_srvo3}. These studies highlight the necessity to explicitly account for electron correlation when calculating material properties, beyond those given by basic one-particle LDA. {\srvo} has a perovskite structure with completely occupied oxygen $2p$ bands, and partially occupied vanadium $3d$ bands.

\subsection{2. Calculation of SrVO$_{3}$ Green's functions and self-energies}

A basic electronic structure calculation for {\srvo} was carried out in CASTEP.
{\srvo} has a perovskite unit cell (space group of crystal = 221: Pm-3m, -P 4 2 3) with lattice parameters $a=b=c=3.8421${\AA} \cite{amadon_srvo3}, giving a unit cell volume $V = 56.72${\AA}$^{3}$. We have used a $20$ x $20$ x $20$ Monkhorst-Pack $k$-point grid, with $550$ irreducible $k$-points. The pseudopotentials for all three elements $Sr$, $V$ and $O$ were taken from the C17 CASTEP set, and the plane-wave basis cut-off was automatically determined to be $653.07$ eV. The calculations were performed at a temperature $T=0.1$ eV ($\beta = 10$ eV$^{-1}$). 

Figure 1 shows the LDA band structure and DoS calculation for {\srvo} obtained from CASTEP. The DoS for $O(2p)$ and $V(3d)$ orbitals are shown, along with the total DoS. There is an isolated set of three partially occupied bands around the Fermi level, which originate mainly from the triply degenerate vanadium $t_{2g}$ orbitals, $(d_{xy}, d_{xz}, d_{yz})$. The contribution from $V(e_{g})$ and $O(2p)$ orbitals is minimal in the vicinity of $\epsilon_{f}$.

\begin{figure}[h]
 \includegraphics[width=0.45\textwidth]{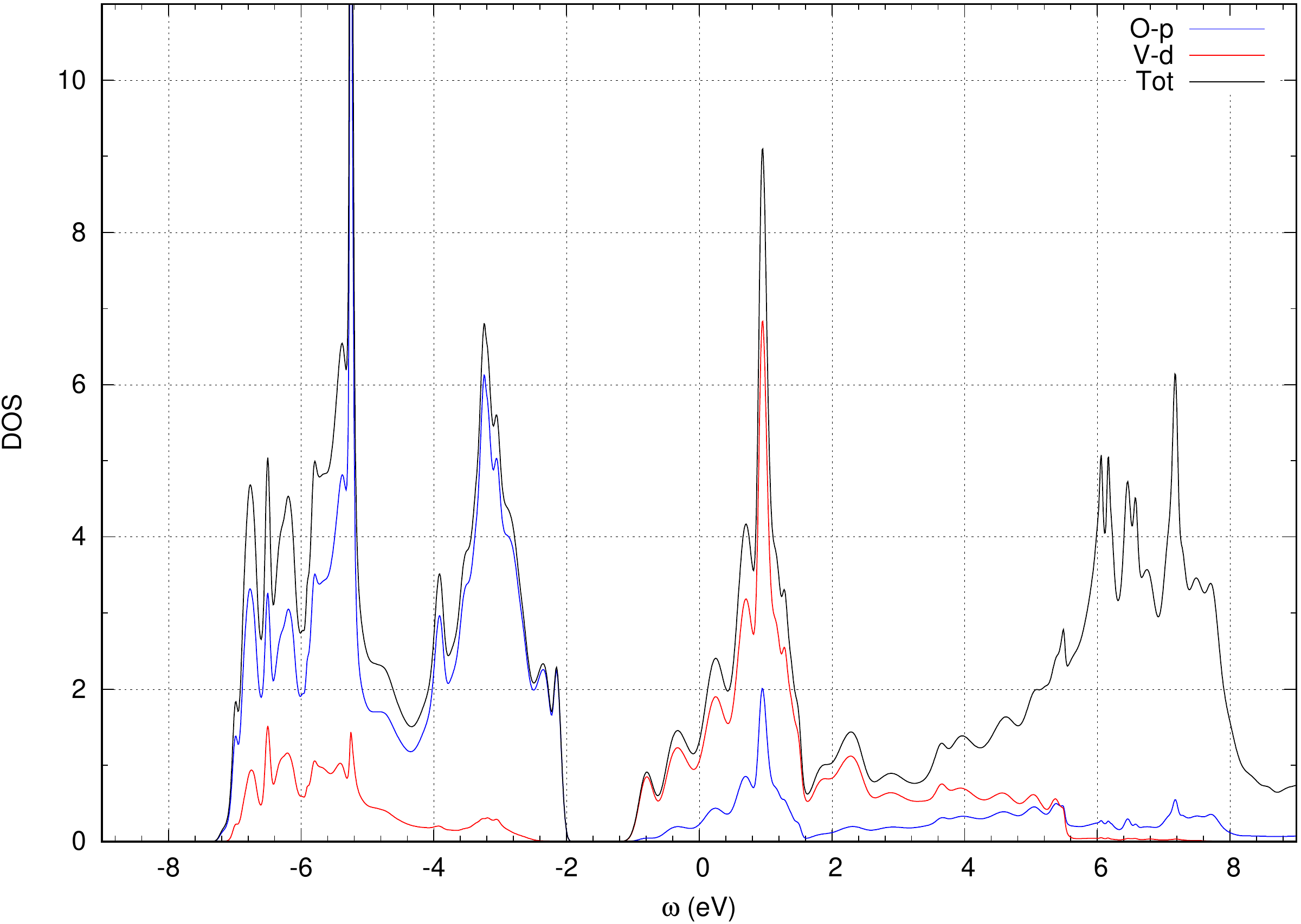}
 \includegraphics[width=0.45\textwidth]{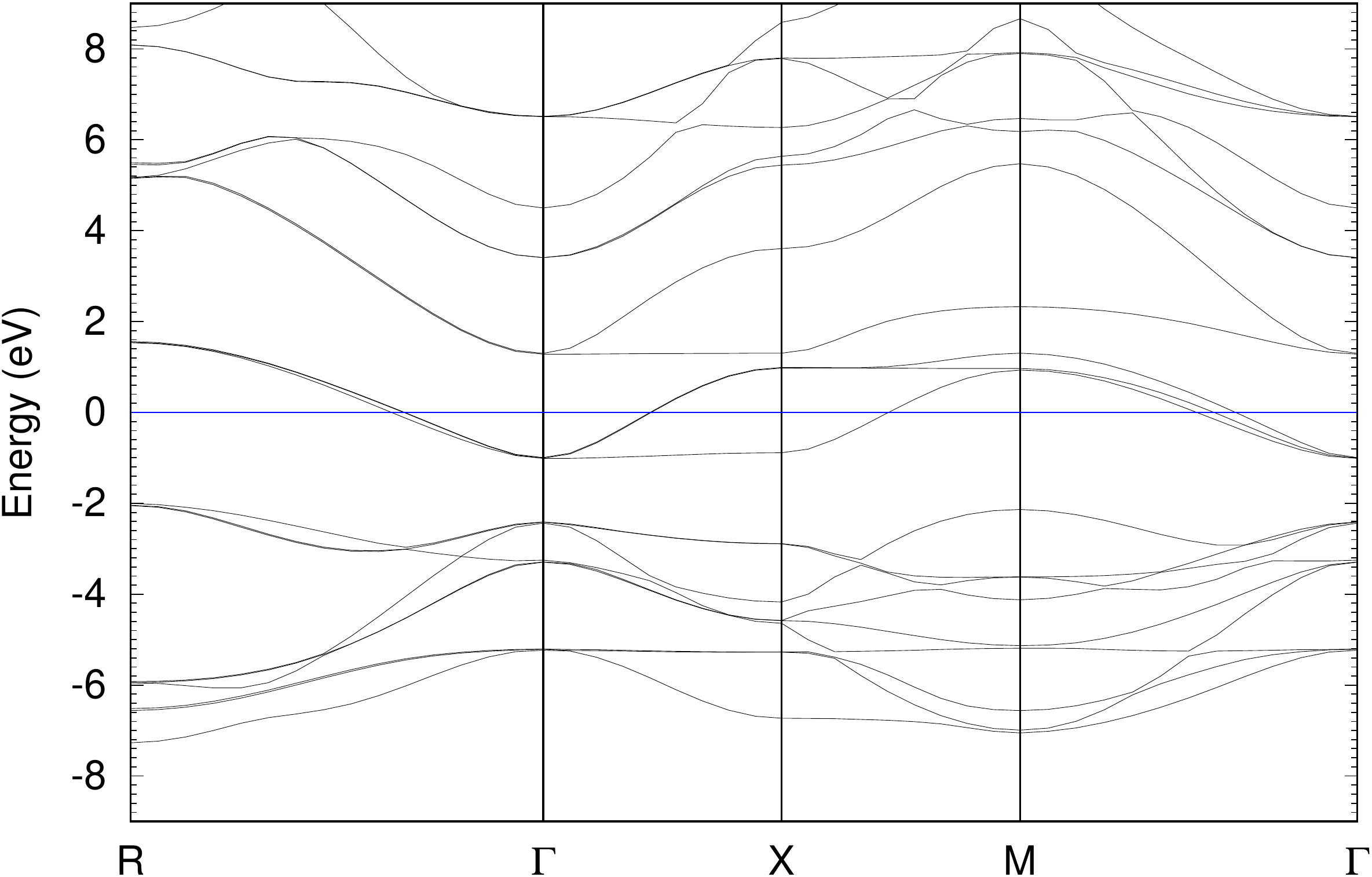}
  \label{fig:Fig_1}
  \caption{   (Upper panel)   The corresponding metallic DoS for perovskite SrVO$_{3}$ from CASTEP LDA.. (Lower panel) CASTEP LDA bandstructure for perovskite SrVO$_{3}$. The isolated set of three partially occupied bands around the Fermi Level (blue line) are formed from the almost triply degenerate vanadium $t_{2g}$  orbitals, $(d_{xy}, d_{xz}, d_{yz})$.}
\end{figure}

Particularising our QMC solver to the three vanadium $t_{2g}-d$ orbitals, $M=3$, gives $\mu = 5U/2 - 5J$, and $M(2M-1)=15$ auxiliary fields are required for the multi-orbital quantum Monte Carlo simulation. 
  
Previous studies with ED-like solvers have used $< 5$ bath levels to parameterise the Weiss field \cite{liebsch}, \cite{rost} at inverse temperatures of $\beta \approx 10$. In our calculation we set the number of bath levels $N_{b}=5$; 
in fact, test calculations with $N_{b}$ as great as $9$ showed no significant 
change to the impurity Green's function at this temperature. It should be 
noted that the computational load of the QMC simulation scales as $ \propto N_{b}^{3}\beta$.    

The results of the impurity QMC Green's function are shown in Figure 2a for 
the {\it first} iteration of the DMFT calculation for {\srvo} across the 
interval $0 \leq \tau \leq \beta$. Two QMC Green's functions are shown 
for one of the $t_{2g}-d$ orbitals for imaginary time discretisations $L=50$ 
and $L=80$, along with fits to these functions using the method described 
in Section III. The parameters chosen for this calculation are $U=4$, $J=0$ and $\beta = 10$, and match those used in other DFT+DMFT studies of {\srvo}. The effect of the Trotter error is evident, and in Figure 2b a detailed portion of 
the Green's function is shown, along with the Green's function that 
results from our integrated fitting and extrapolation procedure. In this
example a set of grids ${L = 50, 60, 70, 80}$ were used. However identical
results are also seen when using either ${L = 50, 60, 70}$ or ${L = 60, 70, 80}$, 
for instance. Additionally, a minimum of $N_l = 22$ Legendre polynomials are needed to
correctly express $G(\tau)$ across the interval, but especially so near the boundaries where
its derivative is larger. For comparison, the equivalent TRIQS CT-QMC result is 
also shown, and the agreement between the two results can be readily observed. 
This demonstrates that the QMC solver combined with our interpolation-extrapolation method is capable of generating quasi-continuous time solutions to 
the impurity problem.

\begin{figure}[h]
  \includegraphics[width=0.5\textwidth]{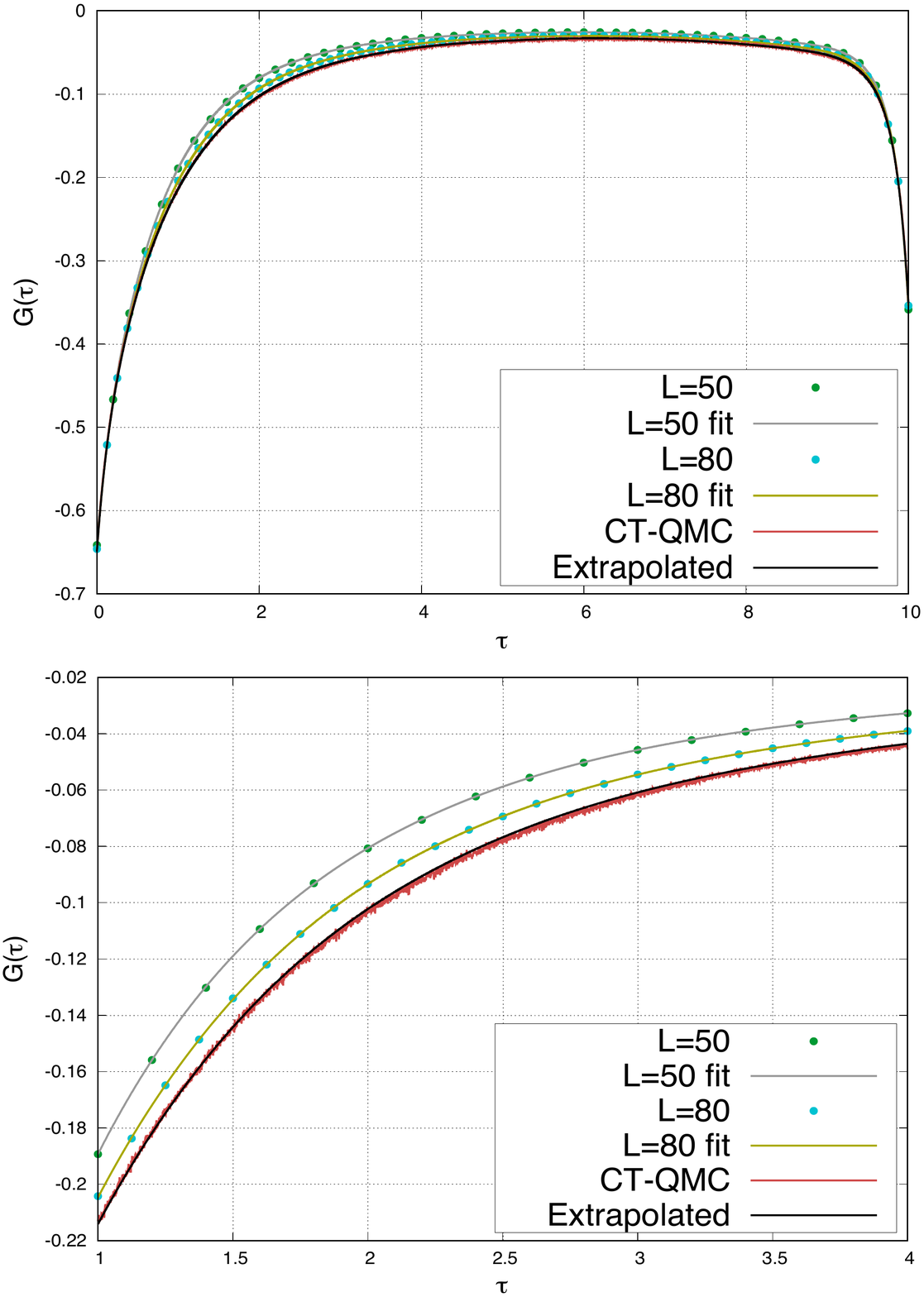}
  \label{fig:Fig_2}
  \caption{   BSS-QMC impurity Green's functions for a selected t$_2g$ SrVO$_{3}$ orbital at $U=4$, $J=0$, $T=0.1$ on successive discrete imaginary time grids for the first DMFT iteration. The extrapolation of the fitted functions, which use  $N_l=22$ Legendre polynomials, to $\Delta \tau \rightarrow 0$  is shown. Also included is the CT-QMC result, which is free of Trotter error. The extrapolation procedure is using the data obtained at respectively $L=50, 60, 70, 80$.
}
\end{figure}

To test the method further, an additional calculation was performed for
$U=8$, $J=0.65$ and $\beta = 10$. Figure 3a shows the Green's functions and
their fits for the same time discretisations as before, illustrating the
enhanced Trotter error for this larger $U$ value. In Figure 3b a comparison
with the extrapolated Green's function with a QMC calculation for $L=200$
is shown. It is expected that the quasi-continuous solution will be very
close to the very fine-scale (but computationally demanding) QMC calculation,
and this is indeed the case.

\begin{figure}[h]
  \includegraphics[width=0.5\textwidth]{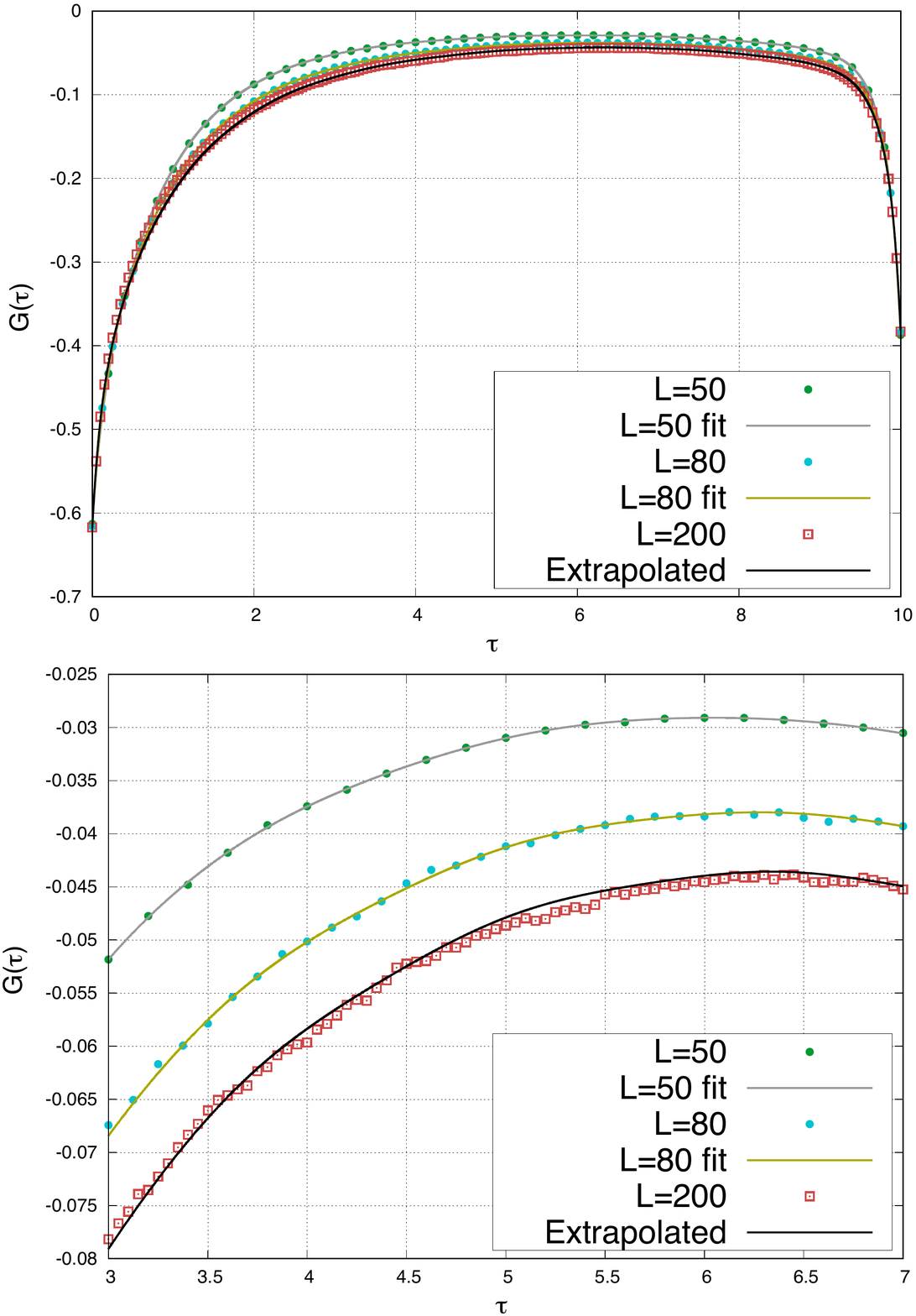}
  \label{fig:Fig_3}
  \caption{    BSS-QMC impurity Green’s functions for a selected t$_2g$ SrVO$_{3}$ orbital at $U=8$, $J=0.65$, $T=0.1$.
}
\end{figure}

The next step is the calculation of the self-energy, which involves Fourier transforming the Green's function from $\tau$ to $i \omega_{n}$ \textit{i.e.}
\begin{equation}
\Sigma(i \omega_{n}) = {\cal{G}}_{0}^{-1}(i \omega_{n}) - G_{QMC}^{-1}(i \omega_{n}) 
\end{equation}
where $G_{QMC}^{-1}$ is the inverse of the calculated quasi-continuous time Green's function. Figure 4 shows the self-energy result for the first DMFT iteration for a 
set of different $L$ values at $U=8$, $J=0.65$, $\beta = 10$ . The extrapolated result, for both the real (Figure 4a) and imaginary (Figure 4b) parts, is a close match to the fine-scale QMC calculation. Clearly, discretisation of the
imaginary time interval has a significant effect of self-energies.

\begin{figure}[h]
  \includegraphics[width=0.5\textwidth]{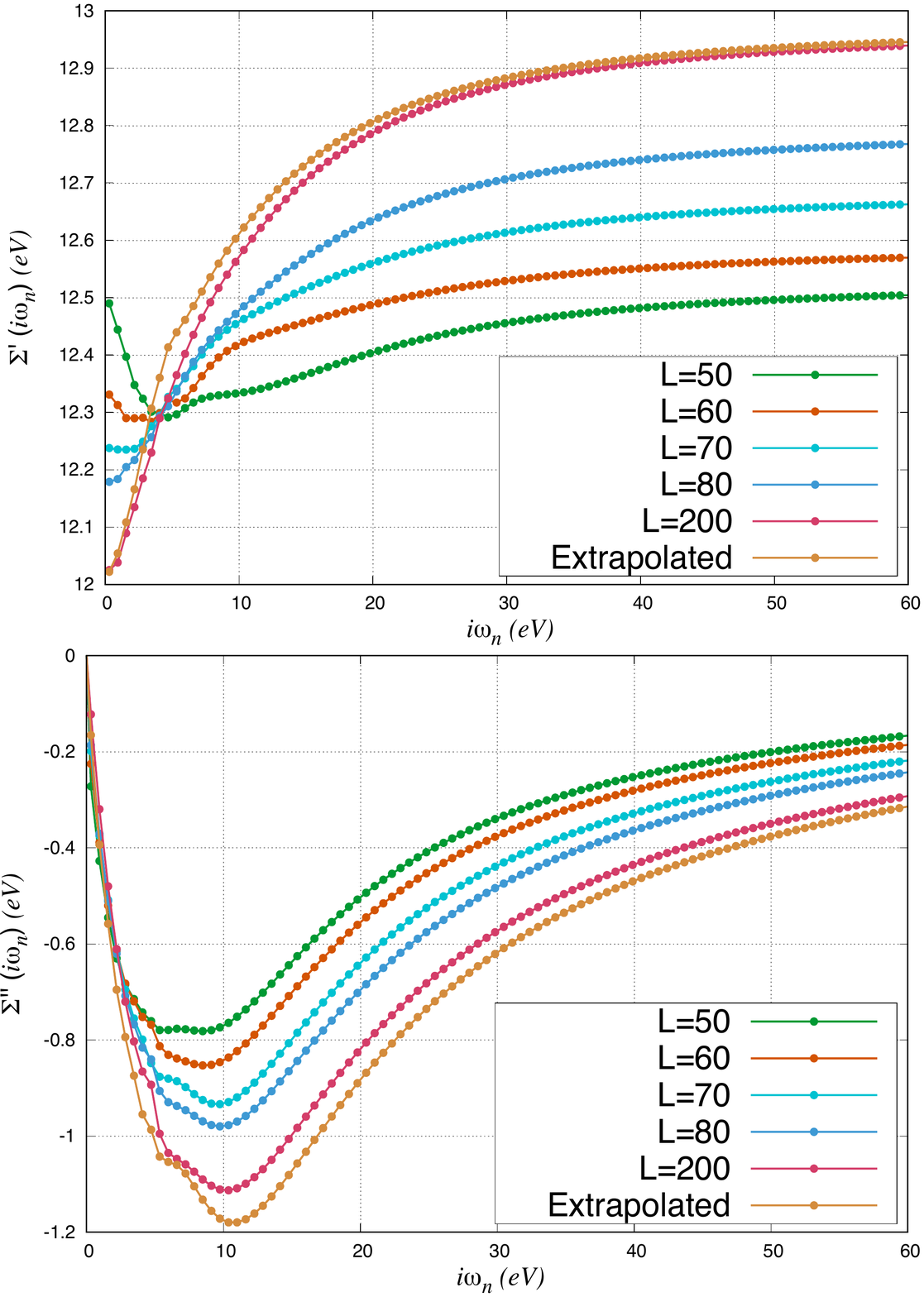}
  \label{fig:Fig_4}
  \caption{   The DMFT self-energy at $U=8$, $J=0.65$ and $T=0.1$ in the limit of $\Delta \tau \rightarrow 0$ for the first DMFT iteration. The excellent agreement of the extrapolated data in both real and imaginary parts with the $L=200$ result is
a consequence of the systematic nesting of the coarse-grid set of discrete BSS-QMC results, their fitting and subsequent extrapolation.
}
\end{figure}

In Figure 5 the self-energy calculations are now taken to full DMFT convergence. 
It is not computationally feasible to perform a fully converged self-consistent 
calculation using an imaginary time discretisation of $L=200$. As justified in 
Figure 4, by extrapolating a finite set of coarse discrete time grids it is,
however, possible to replicate the continuous result with this method. 
Therefore, it is possible to converge the calculation using a subset of 
successively coarse grids that achieve the same accuracy as that of the 
more expensive fine grid calculation. We see that after three iterations 
the DMFT converges, as SrVO3 is relatively weakly correlated.

\begin{figure}[h]
  \includegraphics[width=0.5\textwidth]{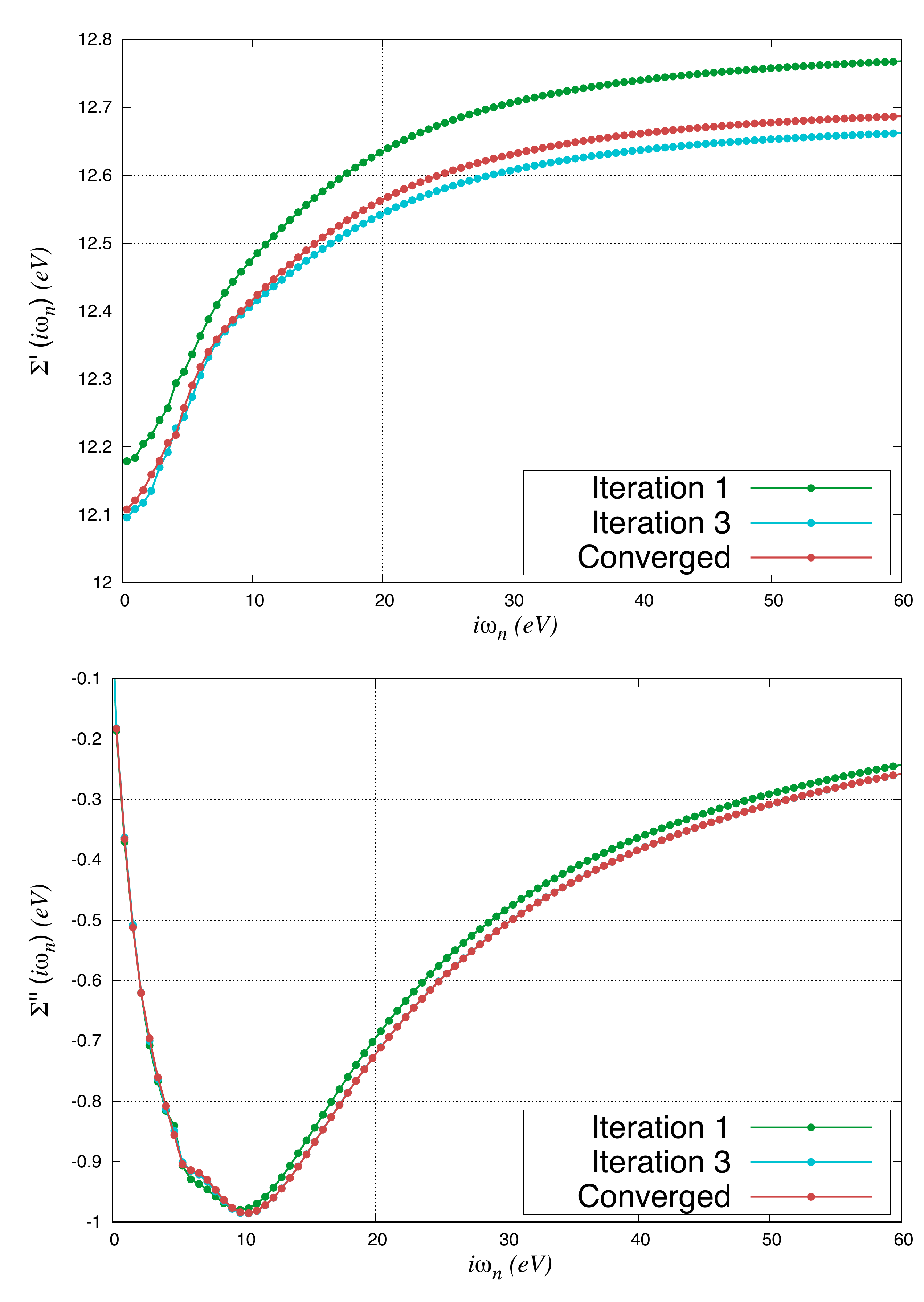}
  \label{fig:Fig_5}
  \caption{   The self-energy during a self consistent DMFT cycle of a selected t$_2g$ SrVO$_{3}$ orbital at $U=8$, $J=0.65$ and $T=0.1$. SrVO$_{3}$ is weakly correlated with a small quasi-particle weight, and it converges after five iterations.
}
\end{figure}

The QMC solver and its associated quasi-continuous time protocol can
be straightforwardly applied to ever-larger values of electron
correlation.  In Figure 6 calculations are shown for $U=12$, $J=0.65$,
$\beta = 10$.  For this large $U$ value the Trotter error on the
Green's function calculation is non-negligible. For one of the
orbitals, Figure 6a shows the raw QMC Green's function (with
interpolations), along with an extrapolation to quasi-continuous
time. Figure 6b shows that at this significant level of correlation the
Trotter error is dramatically enhanced in contrast to smaller
values of $U$. As a result, it is necessary to extrapolate with 
a finer set of discrete grids, i.e L= $80, 90, 100$, while also 
increasing the size of the Legendre basis to $N_l=28$  
polynomials. In doing so, it is possible to recover the correct 
form of the extrapolation scaling and therefore remove the non-trivial 
Trotter error induced by such strong interactions. 

\begin{figure}[h]
  \includegraphics[width=0.5\textwidth]{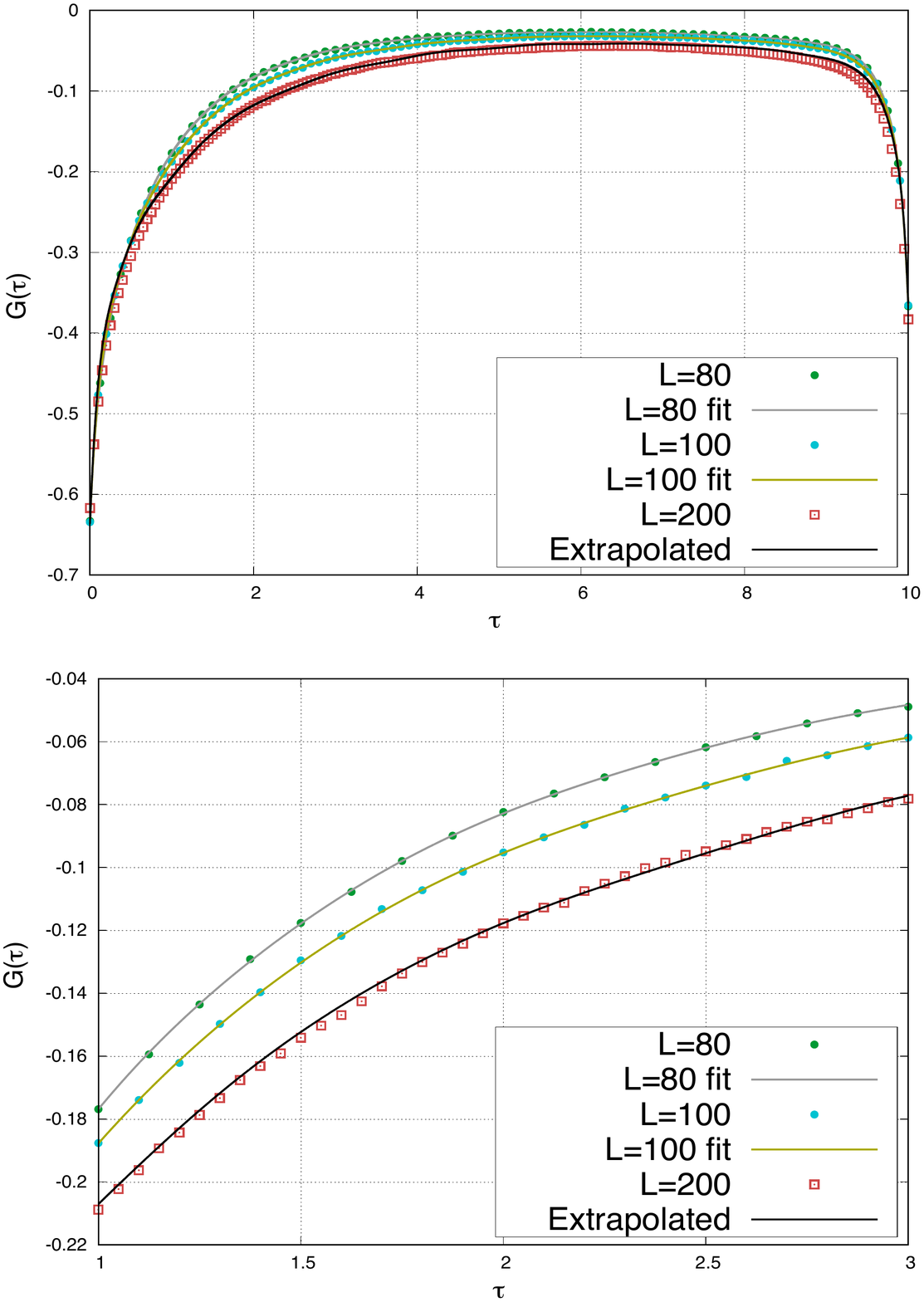}
  \label{fig:Fig_6}
  \caption{  Upper panel: (6a) BSS-QMC impurity Green's function in the strongly interacting limit of $U=12$, $J=0.65$ and $T=0.1$. Lower Panel: (6b) The relative separation of the successive grids is indicative of the severe systematic Trotter error present in BSS-QMC calculations for large values of $U$. Using the quasi-continuous orthogonal polynomial method, by extrapolating respectively the $L=80, 90, 100$ grids with  $N_l=28$ Legendre polynomials, illustrates how it can be remedied by comparing with the approximately continuous $L=200$ data.}
\end{figure}

%%%%%%%%%%%%%%%%%%%% END OF SECTION IV %%%%%%%%%%%%%%%%%

\section{V. Conclusions}

In recent years the DFT+DMFT framework has emerged as a powerful and effective procedure for undertaking \textit{ab-initio} calculations of the properties of real materials when electron correlation effects are a significant influence. The method builds on the well-established capabilities of DFT, whilst seeking to enhance the way the electron correlation problem is addressed by utilising a quantum many-body physics approach. 

A big step towards demonstrating the full potential of the DFT+DMFT scheme would be to attempt fully charge self-consistent calculations for many more materials. However, to do this requires computationally efficient and accurate quantum impurity solvers that work over a very broad range of parameters. The work presented in this paper presents one possible implementation of a QMC-based solver that addresses this need. Additionally, applications such as the calculation of material equations of 
state, where many repeated DFT+DMFT calculations are needed, also require very fast and accurate QMC solvers. The solver we have presented here has the advantage of scaling linearly in inverse temperature, and moreover provides a quasi-continuous imaginary time solution. We have integrated the solver with the popular DFT code CASTEP, thus opening the way to fully charge self-consistent calculations on real materials. Though the validation of this approach has been done here against {\srvo} it will be of particular value when used in the more challenging context of modelling the properties of $f$-band materials, and particularly for calculating materials' equations of state.

%%%%%%%%%%%%%%%%%%%% END OF SECTION V %%%%%%%%%%%%%%%%%% 

\section{Acknowledgements}

The authors thank the EPSRC for support through grant EP/M011038/1, and AWE for support through its Future Technologies fund. This work used the ARCHER UK National Supercomputing Service, for which access was obtained via the UKCP consortium, and funded by EPSRC grant EP/P022561/1.

%%%%%%%%%%%%%%% END OF ACKNOWLEDGEMENTS %%%%%%%%%%%%%%%%

\section{Appendices}

\subsection{Appendix A: Updating the chemical potential}
During each cycle in the DMFT iteration it is necessary to adjust the chemical potential in the Bloch Green's function (equation (\ref{eq:gf_bloch})) in order to maintain the overall correct number of electrons. Essentially, at each iteration a value of $\mu$ is sought that ensures the constancy of the overall electron occupancy, equation (\ref{eq:occ_bloch}). A search algorithm employing the Brent method \cite{numerical_recipes} is used to fix $\mu$.   

\subsection{Appendix B: Double-counting correction}
CASTEP calculations already include a partial representation of electron correlation effects through the DFT exchange-correlation potential. In order not to count the contribution due to electron correlation twice the first step is to subtract the effect of the DFT potential through the ``double-counting'' approximation.

Double counting is not unambiguously resolved in DFT+DMFT, and there is no incontestable way to perform this correction. In CASTEP there is a choice of three double counting corrections: i) Fully localised limit (FLL); ii) Around mean-field limit, and iii) Held's correction \cite{DMFT_review} \cite{held_correction} \cite{pavarini_julich},\cite{evgeny_Ce}. A selection can be made according to the modelling context. Taking each in turn:

i) FLL: In this approximation it is assumed that an orbital occupation is either $0$ or $1$ \textit{i.e.} empty or full. Denoting $N_{\sigma} = \sum_{m} n_{m\sigma}$, $N_{tot}=\sum_{\sigma}N_{\sigma}$ and taking an average value for $U$ and $J$ as follows 
% no equation numbering in appendix
\[ U_{avg} = U = \frac{1}{(2l+1)^{2}}\sum_{m, m'}U_{mm'} \]
and
\[ U_{avg} - J_{avg} = U - J = \frac{1}{2l(l+1)}\sum_{m,m'}(U_{mm'} - J_{mm'}),    \]
the double counting is found to be
\[ E_{DC} = \frac{1}{2}UN_{tot}(N_{tot}-1)-\frac{1}{2}J\sum_{\sigma}N_{\sigma}(N_{\sigma}-1).\]
Differentiating with respect to $N_{\sigma}$ gives
\[V_{\sigma}^{DC} = U\left(N_{tot}-\frac{1}{2}\right) - J\left(N_{\sigma}-\frac{1}{2}\right).\]
This approximation is better suited to insulating systems.

ii) AMF: Here it is assumed that the average occupation of an orbital ($n_{m\sigma}$) is independent of $m$, so that
\[ n_{m\sigma}=n_{\sigma}\equiv\frac{N_{\sigma}}{2l+1} \]
where $N_{\sigma}$ is the total occupation of the impurity site (with spin $\sigma$) and $l$ orbitals. The double-counting energy is given by 
\[ E_{DC} = Un_{\uparrow}n_{\downarrow} + \frac{2l}{2l+1}\frac{U-J}{2}(n_{\uparrow}^{2} + n_{\downarrow}^{2})         \]
and the potential by
\[ V_{DC} = U\left(N_{tot} - \frac{n_{\sigma}}{2l+1}\right) - J\left(n_{\sigma} - \frac{n_{\sigma}}{2l+1}\right).      \] 
This approximation is better suited to metallic systems.

iii) Held's correction: In this approximation an average Coulomb repulsion $U_{av}$ is introduced as follows 
\[ U_{av} = \frac{U + (l-1)(U-2J) +(l-1)(U-3J)}{2l-1} \]
where $l$ is the degeneracy of the shell. The double counting and associated potential as given as
\[ E_{DC} = \frac{U_{av} N_{tot}(N_{tot}-1)}{2} \] 
and
\[ V_{\sigma}^{DC} = U_{av} \left(N_{tot}-\frac{1}{2}\right). \]

%%%%%%%%%%%%%%%%%%%% END OF APPENDICES %%%%%%%%%%%%%%%%%% 

%%%%%%%%%%%%%%%%%%%%%%%%%%%%%%%%%%%%%%%%%%%%%%%%%%%%%%%%%
%%%%%%%%%%%%%%%%%%%%%%%%%%%%%%%%%%%%%%%%%%%%%%%%%%%%%%%%%

$^{*}$ christopher.rhodes@kcl.ac.uk

\bibliographystyle{prsty}

\begin{thebibliography}{10}



\bibitem{CT-QMC_1}
A. N. Rubtsov, V. V. Savkin and A. I. Lichtenstein, Continuous-time quantum Monte Carlo method for fermions. Physical Review B {\bf 72}, 035122 (2005).

\bibitem{CT-QMC_2}
P. Werner, A. Comanac, L. de Medici, M. Troyer and A. J. Millis, Continuous-Time Solver for Quantum Impurity Models. Physical Review Letters {\bf 97}, 076405 (2006).

\bibitem{CT-QMC_3}
P. Werner and A. J. Millis, Hybridization expansion impurity solver: General formulation and application to Kondo lattice and two-orbital models. Physical Review B {\bf 74}, 155107 (2006).

\bibitem{CT-QMC_4}
E. Gull, A. J. Millis, A. I. Lichtenstein, A. N. Rubtsov, M. Troyer and P. Werner, Continuous-time Monte Carlo methods for quantum impurity models. Reviews of Modern Physics {\bf 83}, 349 (2011).

\bibitem{khatami}
E. Khatami, C. R. Lee, Z. J. Bai, R. T. Scalettar and M. Jarrell, Cluster solver for dynamical mean-field theory with linear scaling in inverse temperature. Physical Review B {\bf 81}, 056703 (2010).

\bibitem{rost}
D. Rost, F. Assaad and N. Blumer, Quasi-continuous-time impurity solver for the dynamical mean field theory with linear scaling in the inverse temperature. Physical Review E {\bf 87}, 053305 (2013).

\bibitem{CASTEP}
S. J. Clark, M. D. Segall, C. J. Pickard, P. J. Hasnip, M. J. Probert, K. Refson and M. C. Payne, First principles methods using CASTEP", Zeitschrift fuer Kristallographie {\bf 220}(5-6), 567-570 (2005).  

\bibitem{DFT-ref}
Materials Modelling using Density Functional Theory: Properties and Predictions. F. Giustino. Oxford University Press, Oxford, 2014.

\bibitem{GW+DMFT}
J. M. Tomczak, P. Liu, A. Toschi, G. Kresse and K. Held, Merging GW with DMFT and non-local correlations beyond. European Physics Journal (Special Topics) {\bf 226}, 2565–2590(2017). 

\bibitem{DMFT_review}
A. Georges, G. Kotliar, W. Krauth and M. J. Rozenberg, Dynamical mean-field theory of strongly correlated fermion systems and the limit of infinite dimensions. Reviews of Modern Physics {\bf 68}, 13 (1996). 

\bibitem{CT-QMC_linear}
M. Iazzi and M. Troyer, Efficient continuous-time quantum Monte Carlo algorithm for fermionic lattice models. Physical Review B {\bf 91}, 241118(R) (2015).

\bibitem{mehta}
S. Mehta, G. D. Price and D. Alf\'e, \textit{Ab-initio} thermodynamics and phase diagram of solid magnesium: A comparison of the LDA and GGA. The Journal of Chemical Physics {\bf 125}, 194507 (2006). 

\bibitem{erik}
E. R. Ylvisaker, DFT and DMFT: Implementations and Applications to the Study
of Correlated Materials. PhD Thesis, University of California Davis (2008). 

\bibitem{mcmahan}
A. K. McMahan, Combined local-density and dynamical mean field theory calculations for the compressed lanthanides Ce, Pr and Nd. Physical Review B {\bf 72}, 115125 (2005).

\bibitem{evgeny_Ce}
E. Plekhanov, P. Hasnip, V. Sacksteder, M. Probert, S. J. Clark, K. Refson and C. Weber, Many-body renormalisation of forces in \textit{f}-materials. Physical Review {\bf B} 98, 075129 (2018).  

\bibitem{dos_santos}
R. R. dos Santos, Introduction to Monte Carlo simulations for fermionic systems. Brazilian Journal of Physics {\bf 33}, 36 (2003).

\bibitem{caffarel}
M. Caffarel and W. Krauth, Exact diagonalisation approach to correlated fermions in infinite dimensions: Mott transition and superconductivity. Physical Review Letters {\bf 72}, 1545 (1994). 

\bibitem{cedric_ed}
C. Weber, A. Americci, M. Capone and P. B. Littlewood, Augmented hybrid exact-diagonalization solver for dynamical mean field theory. Physical Review B {\bf 86}, 115136 (2012). 

\bibitem{sakai}
S. Sakai, R. Arita, K. Held and H. Aoki, Quantum Monte Carlo study for multiorbital systems with preserved spin and orbital rotational symmetries. Physical Review B {\bf 74}, 155102 (2006).

\bibitem{sakai_2004}
S. Sakai, R. Arita, and H. Aoki, Numerical algorithm for the double-orbital Hubbard model: Hund-coupled pairing symmetry in the doped case. Physical Review B {\bf 70}, 172504 (2004).

\bibitem{han}
J.E. Han, Spin-triplet s-wave local pairing induced by Hund’s rule coupling. Physical Review B {\bf 70}, 054513 (2004).

\bibitem{motome_I}
Y. Motome and M. Imada, A Quantum Monte Carlo Method and Its Applications to Multi-Orbital Hubbard Models. Journal of the Physical Society of Japan {\bf 66}, 1872 (1997).

\bibitem{motome_II}
Y. Motome and M. Imada, Numerical Study for the Ground State of Multi-Orbital Hubbard Models. Journal of the Physical Society of Japan {\bf 67}, 3199 (1998).

\bibitem{held_vollhardt}
K. Held and D. Vollhardt, Microscopic conditions favoring itinerant ferromagnetism: Hund's rule coupling and orbital degeneracy. European Physics Journal B {\bf 5}, 473 (1998).

\bibitem{BSS}
R. Blankenbecler, D. J. Scalapino and R. L. Sugar, Monte Carlo calculations of coupled boson-fermion systems I. Physical Review D {\bf 24}, 2278 (1981). 

\bibitem{white}
S. R. White, D. J. Scalapino, R. L. Sugar, E. Y. Loh, J. E. Gubernatis, and R. T. Scalettar, Numerical study of the two-dimensional Hubbard model. Physical Review B {\bf 40}, 506 (1989).

\bibitem{loh+gub}
E. Y. Loh and J. E. Gubernatis, Stable numerical simulations of models of interacting electrons in condensed matter physics. Chapter 4 in Electronic Phase Transitions \textit{ed.} W. Hanke and Yu.V. Kopaev. Elsevier Science Publications (1992).

\bibitem{vidberg}
H. J. Vidberg and J. W. Serene, Solving the Eliashberg equations by means of N-point Pad\'e approximants. Journal of Low Temperature Physics {\bf 29}, 179 (1977).

\bibitem{nukala}
P. K. V. V. Nukala, T. A. Maier, M. S. Summers, G. Alvarez, and T. C. Shulthess,
Fast update algorithm for the quantum Monte Carlo simulation of the Hubbard model.
Physical Review B {\bf 80}, 195111 (2009).

\bibitem{KPM_ref}
A. Wei\ss e, G. Wellein, A. Alvermann and H. Fehske, The kernel polynomial method. Reviews of Modern Physics {\bf 78}, 275 (2006).

\bibitem{Green_Lagrange}
L. Boehnke, H. Hafermann, M. Ferrero, F. Lechermann, and O. Parcollet, Orthogonal polynomial representation of imaginary time Green's functions. Physical 
Review B {\bf 84}, 075145 (2011).

\bibitem{anisimov_srvo3}
V. I. Anisimov, D. E. Kondakov, A. V. Kozhevnikov, I. A. Nekrasov, Z. V. Pchelkina, J. W. Allen, S.-K. Mo, H.-D. Kim, P. Metcalf, S. Suga, A. Sekiyama, G. Keller, I. Leonov, X. Ren, and D. Vollhardt, Full orbital calculation scheme for materials with
strongly correlated electrons. Physical Review B {\bf 71}, 125119 (2005).

\bibitem{nekrasov_srvo3}
I. A. Nekrasov, G. Keller, D. E. Kondakov, A. V. Kozhevnikov, T. Pruschke, K. Held, D. Vollhardt and V. I. Anisimov, Comparative study of correlation effects in CaVO$_{3}$ and {\srvo}. Physical Review B {\bf 72}, 155106 (2005).

\bibitem{lechermann_srvo3}
F. Lechermann, A. Georges, A. Poteryaev, S. Biermann, M. Posternak, A. Yamasaki, and O. K. Anderson, Dynamical mean-field theory using Wannier functions: A flexible route to electronic structure calculations of strongly correlated materials. Physical Review B {\bf 74}, 125120 (2006).

\bibitem{amadon_srvo3}
B. Amadon, F. Lechermann, A. Georges, F. Jollet, T. O. Wehling, and A. I. Lichtenstein, Plane-wave based electronic structure calculations for correlated materials using dynamical mean-field theory and projected local orbitals. Physical Review B {\bf 77}, 205112 (2008).

\bibitem{liebsch}
A. Liebsch and H. Ishada, Temperature and bath size in exact diagonalisation dynamical mean-field theory. Journal of Physics, Condensed Matter {\bf 24}, 053201 (2012).

\bibitem{numerical_recipes}
W. H. Press, B. P. Flannery, S. A. Teukolsky and W. T. Vetterling, Numerical Recipes: The Art of Scientific Computing. Cambridge University Press, 2 edition, (1992).

\bibitem{held_correction}
K. Held, Electronic structure calculations using dynamical mean field theory. Advances in Physics {\bf 56}, 829 (2007).

\bibitem{pavarini_julich}
E. Pavarini, Mott transition: DFT+U vs DFT+DMF. Autumn School on Correlated Electrons: The Physics of Correlated Insulators, Metals, and Superconductors. Forschungszentrum J\"ulich (2017).


  
  


\end{thebibliography}

\end{document}